\documentclass[12pt]{article}
\usepackage{float}
\usepackage[utf8]{inputenc}
\usepackage{amsmath, amssymb}
\usepackage{graphicx}
\usepackage{float}
\usepackage{cite}
\usepackage{hyperref}
\usepackage{authblk}
\usepackage{geometry}
\usepackage{subcaption}
\usepackage{booktabs} 
\usepackage{caption}
\usepackage{xcolor}

\title{\textbf{A NEAT Approach to Evolving Neural-Network-based Optimization of Chiral Photonic Metasurfaces: \\
Application of a Neuro-Evolution Pipeline}}

\author[1]{Davide Filippozzi}
\author[1]{Arash Rahimi-Iman}
\affil[1]{I. Physikalisches Institut and Center for Materials Research, Justus-Liebig-Universität Gießen, D-35392 Giessen, Germany}

\date{\today}

\begin{document}

\maketitle

\begin{abstract}
The design of chiral metasurfaces with tailored optical properties remains a central challenge in nanophotonics due to the highly nonlinear relationship between geometry and chiroptical response. Machine-learning-assisted optimization pipelines have recently emerged as efficient tools to accelerate this process, yet their performance strongly depends on the choice of neural-network (NN) architecture. In this work, we integrate the NeuroEvolution-of-Augmenting-Topologies (NEAT) algorithm into an established deep-learning optimization framework for dielectric chiral metasurfaces. NEAT autonomously evolves both network topology and connection weights, enabling task-specific architectures without manual tuning, whereas the reinforcement-learning strategy in our framework evolves knowledge of the solution space and fine-tunes a model's weights in parallel. Using a pipeline-produced dataset of 9,600 simulated GaP metasurface geometries, we evaluate NEAT under varying input dimensionalities, feature-scaling methods, and data sizes. With standardized feature scaling yielding the most consistent performance for both examined output dimensionalities, the relatively compact NEAT-evolved NN models, when integrated into the full optimization pipeline, achieve similar or improved predictive accuracy and generalization compared to initially employed dense few-layer perceptrons. Accordingly, these resource-efficient models successfully perform inference of metasurfaces exhibiting strong circular dichroism in the visible spectrum, allowing for transfer learning between simulated and experimental data. This approach demonstrates a scalable path toward adaptive, self-configuring machine-learning frameworks for automated photonic design both standalone and as building block for agentic artificial intelligence (AI).
\end{abstract}

\section{Introduction}

Chiral nanophotonic structures and metasurfaces have attracted intense attention due to their capability to tailor light--matter interactions with polarization sensitivity, enabling applications in biosensing, quantum optics, and polarization-resolved imaging~\cite{zhang2022enhancing,solntsev2021metasurfaces,martinez2017single}. The ability of chiral metasurfaces to discriminate between left- and right-circularly polarized light through circular dichroism (CD) and optical activity provides a versatile platform for achieving functionalities that are difficult to realize with conventional photonic systems. However, designing such structures remains a formidable challenge. The optical response of chiral metasurfaces arises from complex electromagnetic coupling effects that depend sensitively on the geometry, arrangement, and material composition of their meta-atoms~\cite{zhu2018giant}. Consequently, forward prediction or inverse design for a target chiroptical response requires advanced optimization strategies capable of exploring a vast, highly non-linear parameter space.

Computational approaches that integrate photonic simulations with machine learning have emerged in recent years and accelerated the design of complex optical structures. Neural networks (NNs) and other data-driven models have demonstrated the capability to "learn" or map the complex relationship between geometrical parameters and optical spectra, offering a pathway toward efficient forward prediction and inverse design~\cite{li2022deep,wang2020deep,elsawy2020numerical,Jiang2019,ma2019probabilistic,deng2021neural,minkov2020inverse,hughes2019forward,Wiecha_2021,li2022empowering}. 

Based on our previous work~\cite{mey2022machine}, we improved a reinforced deep-learning optimization pipeline combining rigorous coupled-wave analysis (RCWA)\cite{Khepri} simulations of spectral properties for in-plane periodic unit-cells with deep learning to iteratively advance prediction capabilities with increased knowledge of the solution space and thereby achieve improvement of the chiroptical response of dielectric metasurfaces. This framework demonstrated that NN-assisted structure--response optimization can be performed fast and with high design accuracy while outperforming simple evolutionary algorithms dedicated to an initial solution finding process (local optimum sampling). Moreover, our recent comparative study addressing the use of the NN pipeline in conjunction with a sophisticated genetic algorithm---utilizing the optimization algorithm developed in Ref.~\cite{Mayer_SPIE2024}---showed that such pipeline can significantly reduce computational cost while maintaining prediction reliability when optimizing these structures regarding both CD and preferred reflectivity simultaneously, for two material configurations and different design complexity~\cite{Filippozzi}. In comparison to the one-value case~\cite{mey2022machine}, the two-feature optimization was pursued in an effort to obtain structures that exhibit high CD without sacrificing the strength of reflectivity in the targeted range~\cite{Filippozzi}, where the NN pipeline overall proved itself as a powerful tool for the fast and effective prototyping/creation of metasurface designs toward specific optical functionalities.

Despite these advances, the provision of an optimal NN architecture as the core model of the optimization algorithm remains a non-trivial task, since the model design strongly influences its performance, generalization, and computational efficiency. Conventionally, NN architectures are selected through manual tuning or heuristic search, which can limit the capability or adaptability of the pipeline and introduce user bias. To overcome this limitation, we introduce in this work a neuroevolutionary approach based on the NEAT algorithm~\cite{stanley2002evolving} to autonomously evolve the structure and connectivity of neural networks within the established optimization framework. In our hybrid optimization framework merging evolution and deep learning, NEAT evolves both network topology and connection weights through genetic principles, progressively increasing network complexity as necessary to improve predictive capability. Consecutively, evolved models undergo finetuning when employed as the new core within the inference-oriented pipeline. This principle allows the architecture itself to adapt dynamically to the underlying physical problem rather than relying on pre-defined network configurations. 

By integrating NEAT into our existing optimization pipeline, we enable a fully automated, adaptive framework for the design of chiral metasurfaces. The neuro-evolved models (i.e. the NEAT-optimized neural networks) learn efficient representations of the nonlinear mapping between structural parameters and chiroptical properties, enhancing model accuracy and convergence speed while circumventing the need for human-based model architecture engineering. This inherently makes such a NEAT-enhanced pipeline a powerful building block in automated design procedures, for instance with the help of autonomous agentic frameworks (cf. agentic optimization of metamaterials in Ref.\cite{lu2025agentic}). Applied to the design of all-dielectric metasurfaces optimized for high circular dichroism in the visible spectrum, our NEAT approach demonstrates superior predictive performance and improved exploration of the design space in comparison to previously tested fixed-topology networks. The work presented here evaluates NEAT under varying input dimensionalities, feature-scaling methods, and data sizes, utilizing our pipeline-evolved dataset of 9,600 simulated GaP metasurface geometries for the neuroevolution algorithm.

\begin{figure}[!ht]
    \centering    \includegraphics[width=1\linewidth]{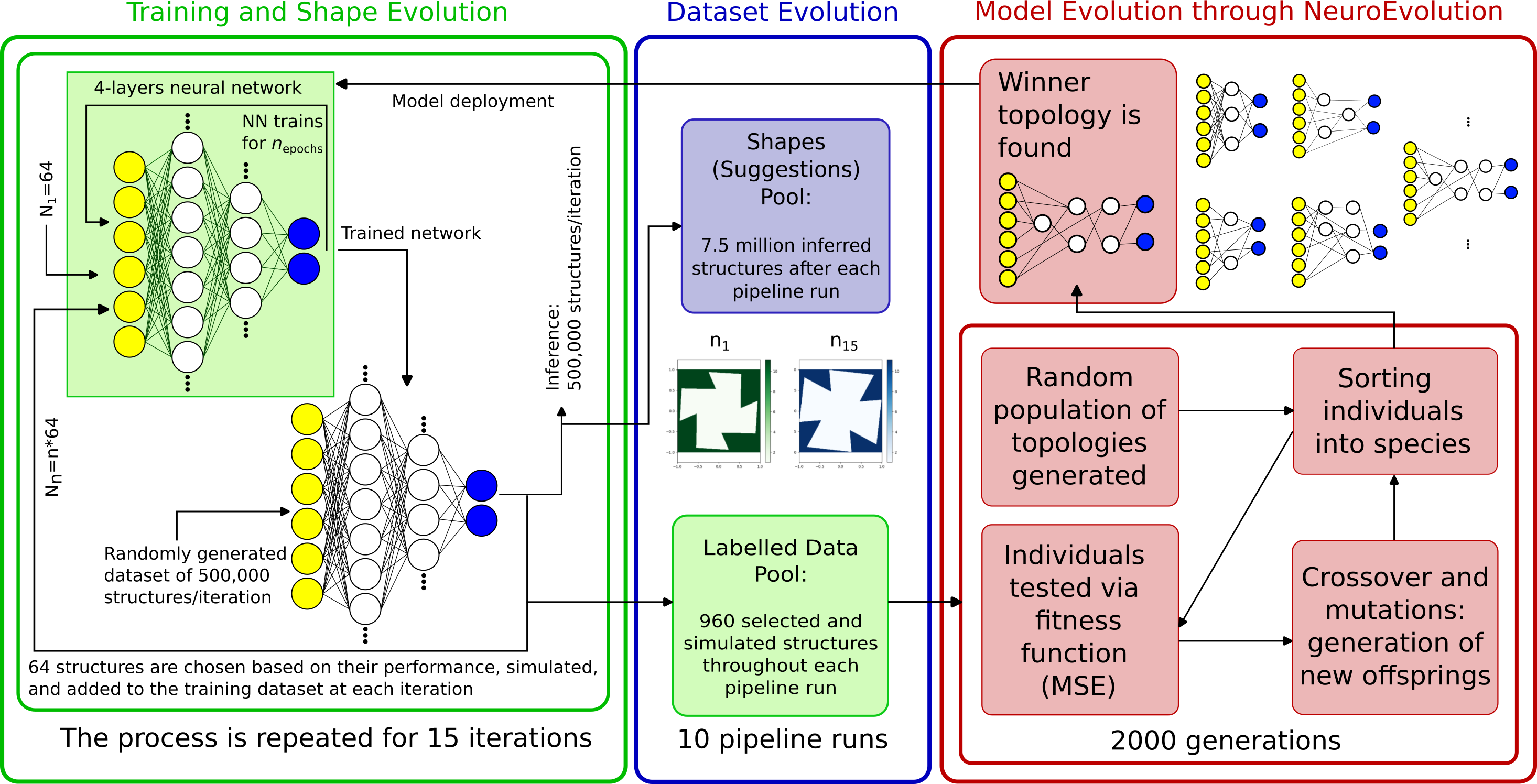}
    \caption{Schematic diagram of the NEAT-enhanced neural-network optimization pipeline. It delivers stochastically a practical data pool (dataset evolution) through iterative inference and simulation steps, evolves structures with the help of the previously reported reinforcement strategy in the deep-learning pipeline (training and shape evolution) and improves the neural network model with the neuroevolution algorithm NEAT (model evolution). As sketched, a few-layer perceptron can act as the initial pipeline model. Green boxes represent the model training part and the acquired labelled data pool for it gradually growing with each pipeline iteration. Example shapes delivered by the pipeline after different iteration steps, here first (greenish) and last (blueish), are indicated centrally. Evolved neural networks via NEAT with six input features (yellow), hidden neurons (white) and two output values (blue nodes) are sketched in the right panel, with an example of a winner model deployed into the prediction pipeline for finetuning indicated.}
    \label{fig:NEATdeepPipeline}
\end{figure}

\section{Methodology}
This work advances our machine learning-assisted optimization pipeline, which we previously established and developed in conjunction with evolution-based approaches for design improvements concerning chiral dielectric metasurfaces~\cite{mey2022machine,Filippozzi}. In the reported frameworks, pre-defined NN models were trained on optical data generated through simulations for a specific optimization task, such as via rigorous coupled-wave analysis (RCWA) to predict the relation between metasurface geometry and chiroptical response, enabling efficient forward design.

Here, we introduce an automated NN architecture discovery step based on the NEAT algorithm. This addition removes the need for manual network design by evolving both the topology and weights of the NN model, thereby adapting the model complexity dynamically to the underlying optimization task.
Figure \ref{fig:NEATdeepPipeline} displays a sketch of our hybrid optimization framework which combines neuro-evolution for model improvement and our reinforced deep-learning algorithm pipeline for efficient structure--response prediction, utilizing iteratively selected feedback via simulation results for capability improvements (the reinforcement strategy used here). The core routine of the algorithm deals with training of the NN model prior to deployment as inference tool. Thereby, it also generates a relevant dataset which can be utilized in the following part for the NEAT-based automated network architecture optimization. 

Once evolved, the resulting network architectures are integrated into the functional NN-based optimization workflow to deliver improvements to the design of chiral metasurfaces within the deep learning building block. The NEAT-generated networks replace the manually designed architectures used in earlier versions of the pipeline, enabling a fully automated process for network topology optimization, model training and structure improvements. The resulting framework provides both predictive and generative capabilities, accelerating the search for geometries exhibiting strong circular dichroism within specified spectral ranges. Ultimately, it enables true automation through the adaptability of the entire optimization pipeline: for a given problem, the model dynamically self-adjusts to the underlying physics.

\subsubsection*{NEAT Implementation}

NEAT was applied to the same optimization problem previously addressed by our fixed-topology neural networks: predicting high circular dichroism ($\Delta R_{CD}$) and preferred handedness reflectance ($R_{pref}$) of circular-polarization sensitive dielectric metasurfaces based on their geometrical parameters. The metasurface unit cell geometry was parameterized by the coordinates of the polygonal corner points defining the meta-atom structure (the unit-cell geometry). Gallium phosphide (GaP) was selected as the material due to its high refractive index and low absorption in the visible spectrum, and the structure thickness was $t=\lambda_0/3$. 

A dataset comprising 9,600 geometries and their corresponding optical responses was generated using the Khepri Python library~\cite{Khepri} for electro-magnetic simulations in multiple runs with different seeds (here 10) of the previously described optimization pipeline for structure selection (see Fig.~\ref{fig:NEATdeepPipeline}). Extracted from the ``Shapes (Suggestions) Pool'' of millions of inferred geometries according to a problem-specific feature function, each selected geometry was simulated under normal incidence for both left- and right-circularly polarized light, and the resulting reflectance spectra were used to compute $\Delta R_{CD}$ and $R_{pref}$. This subset of shapes with CD and reflectivity value labels obtained (indicated in the central part of Fig.~\ref{fig:NEATdeepPipeline} in greenish color as ``Labelled Data Pool'') served as the training and evaluation basis for the NEAT-evolved networks. The result of the NeuroEvolution step (Fig.~\ref{fig:NEATdeepPipeline}, sketched in reddish color to the right) is then inserted into our reinforced supervised learning pipeline (greenish color, left side).

\subsubsection*{Evolution of Neural Networks}

To assess the adaptability and robustness of NEAT for this photonic optimization task, several configurations were tested. 

Regarding dataset size variation, subsets of different sizes were used to investigate the impact of training data quantity on NEAT performance and generalization (10, 50 or 90 percent of the available data). 

In terms of input feature variation, two input sets were compared: (i) the x--y coordinates of three independent corners in the first quadrant (of the unit-cell), and (ii) the x--y coordinates of all twelve corners (that is, the whole unit-cell) generated by symmetry. This allowed testing NEAT's capability for implicit feature selection, i.e., its ability to identify and disregard redundant geometric parameters. Previous application of the few-layer perceptron in our optimization pipeline allowed computationally-efficient training and prediction by exploiting rotation symmetry within the unit cell, using only the three independent corner coordinates in the first quadrant of the structure and keeping the NN model compact~\cite{mey2022machine}. 

Finally, for feature scaling comparison, both normalization and standardization of input and output features were applied to evaluate their influence on training efficiency and convergence. These two feature scaling methods have been chosen as they are amongst the most popular for both NeuroEvolution and standard NN applications.

For each configuration, NEAT was executed with both one or two output neurons, corresponding to only $\Delta R_{CD}$ and $\Delta R_{CD}$ together with $R_{pref}$, respectively. In total, 120 runs were performed, covering all combinations of dataset size, input configuration, and scaling method. Each experiment was repeated five times with different random seeds to ensure statistical validity.

During evolution, NEAT's fitness function was defined as the mean squared error (MSE) between the predicted and simulated optical quantities. The population size, mutation rate, and speciation parameters followed standard settings of the TensorNEAT python library \cite{10.1145/3730406}, and were kept constant across experiments for comparability. The evolution process continued until a maximum generation count was reached (here 2000, cf. Appendix \ref{appendix:hyperparameters} for further details).

In addition, transfer learning of an evolved framework's network trained on simulated data can enable adjustment to experimental data for real metasurfaces, which can in principle link the data-driven design and production recipe improvements directly with the fabrication process and its production outcomes, respectively---ultimately promising automated co-evolution of real-time-data-enhanced models and self-driven fabrication pipelines.

\section{Results and Discussion}

\subsubsection*{Effect of Input Dimensionality and Feature Scaling}
\begin{figure}[!tb]
    \centering
    \includegraphics[width=1\linewidth]{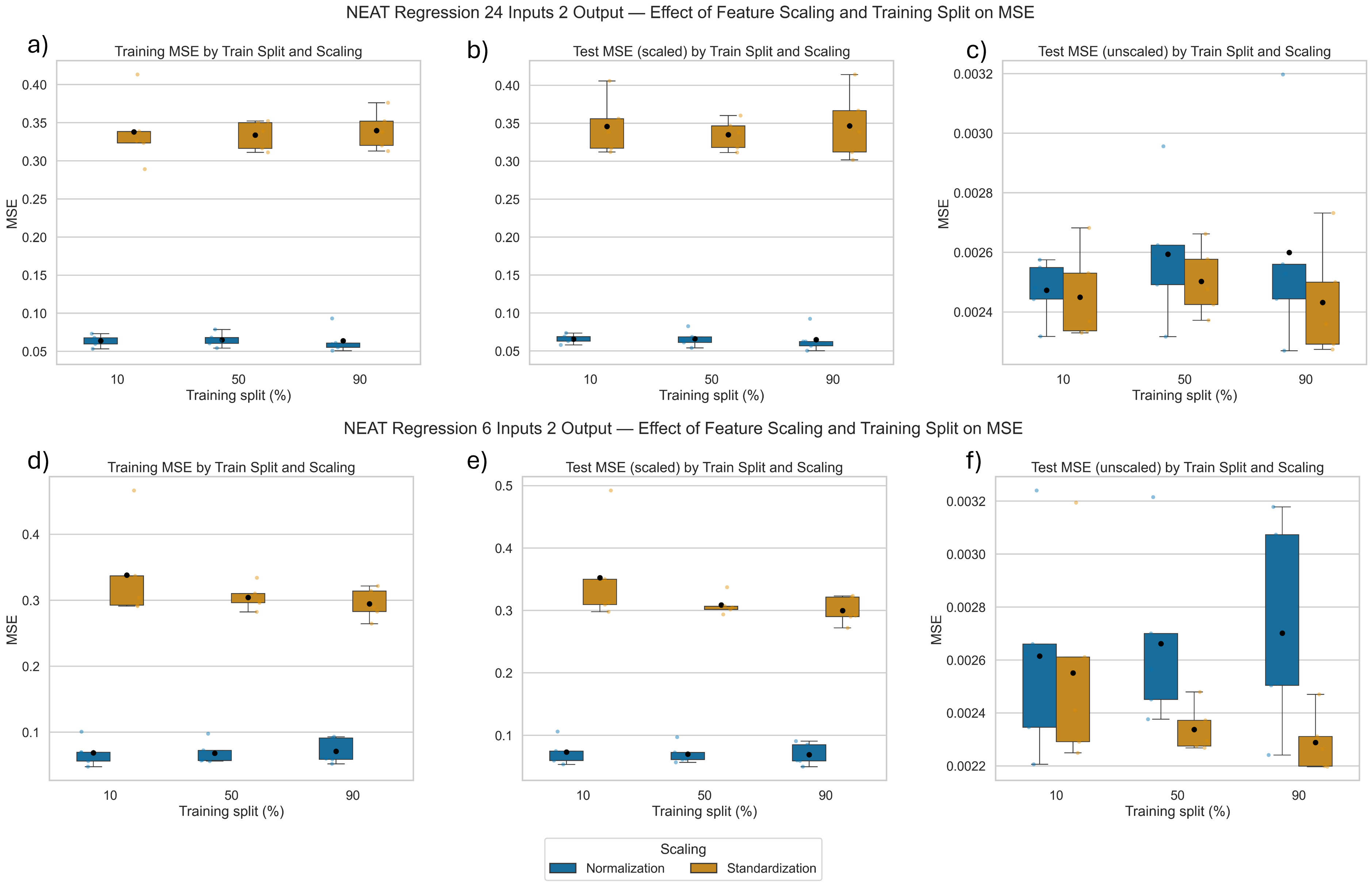}
    \caption{Overview of the NEAT algorithm results obtained for 2 output neurons, with 24 and 6 input neurons used in the model (top and bottom, respectively). a) and d) show the averaged training losses for every train-test split of the dataset in a comparison of normalized (dark colored) and standardized (light colored) data. Error bars obtained from averaging over 5 repetitions are provided for each value. b) and e) similarly show the
    respective averaged test losses. c) and f) summarize the re-scaled results of the
    tests, that is, undoing the "scaler". The direct comparison displays lower MSE values in the case of less input features and standardization for sufficiently sized training data.}
    \label{fig:MSE}
\end{figure}

In a performance comparison between the differently configured NEAT models applied to the 4-fold symmetric chiral unit cell structure, the most accurate models were consistently obtained using six input features combined with standardized feature scaling. This configuration, which utilized the three independent corner coordinates in the first quadrant, yielded the lowest mean squared errors (MSEs) on unseen data, indicating improved generalization compared to networks trained with all 24 symmetry-related inputs.
Figure \ref{fig:MSE} illustrates these results for the two-output configuration predicting circular dichroism ($\Delta R_{CD}$) and preferred handedness reflectance ($R_{pref}$). A two-way ANOVA confirmed that feature scaling significantly influenced network performance for the 6-input case ($p < 0.05$), while dataset size did not have a significant effect on these final \textit{results}. Post hoc analysis (Tukey HSD) further demonstrated that standardization consistently outperformed normalization, reducing the average MSE by approximately 10\%. 

Similar trends are observed for the single-output neuron configurations, the results of which are detailed in the Appendix (Fig.~\ref{fig:NEAT 1 Neurons}).
The errors from NEAT prediction (as is, i.e. before retraining through deployment in the pipeline) for this input/output configuration are about 17\% and of the order of the actual CD value eventually predicted via the pipeline ($\Delta R_{CD} = 0.0094$). While it indicates the reasonable base capabilities provided through NEAT models right away, this also emphasizes the significance of the retraining phase (i.e. weight finetuning through the backpropagation technique) as part of the model deployment in the optimization pipeline, which is described further below.
For the 6-input single-output models, feature scaling remains the dominant factor, with standardization again providing the most robust predictive capabilities. However, for the 24-input configurations, the dataset size exhibits a more significant influence on performance. Specifically, a dataset size of 4,800 training examples (corresponding to a 50\% training split of the 9,600 simulated geometries) provided the lowest MSE.

The observed disparity in performance between these two feature scaling methods, particularly in the 6-input configuration, likely stems from how each technique treats the spatial and statistical integrity of the meta-atom geometry. For the 6-input case, normalization may disrupt the distribution of coordinate points, potentially obscuring critical spatial relationships between the structural parameters and the resulting chiroptical response. One should note that normalizing the coordinates projects them here into a fixed range between -1 and 1, whereas, in contrast to this, standardization centers and scales the data according to the mean value and the variance, respectively. Correspondingly, standardization preserves the underlying correlation between inputs and outputs, which allows the NEAT algorithm to more effectively exploit the dynamic range of the structural features. In the 24-input configuration, the influence of the scaling method is less pronounced because the raw coordinates of the corners are already naturally distributed between -1 and 1 due to the symmetry-based generation of the meta-atom, making the input set inherently similar to a normalized distribution. While standardization modifies the numerical values of the inputs and outputs, its primary advantage remains the preservation of feature correlations, which ensures that the neuroevolutionary process can adaptively capture the nonlinear mapping between geometry and response.

\subsubsection*{Network Topology Evolution}

\begin{figure}[!tb]
    \centering
    \includegraphics[width=1\linewidth]{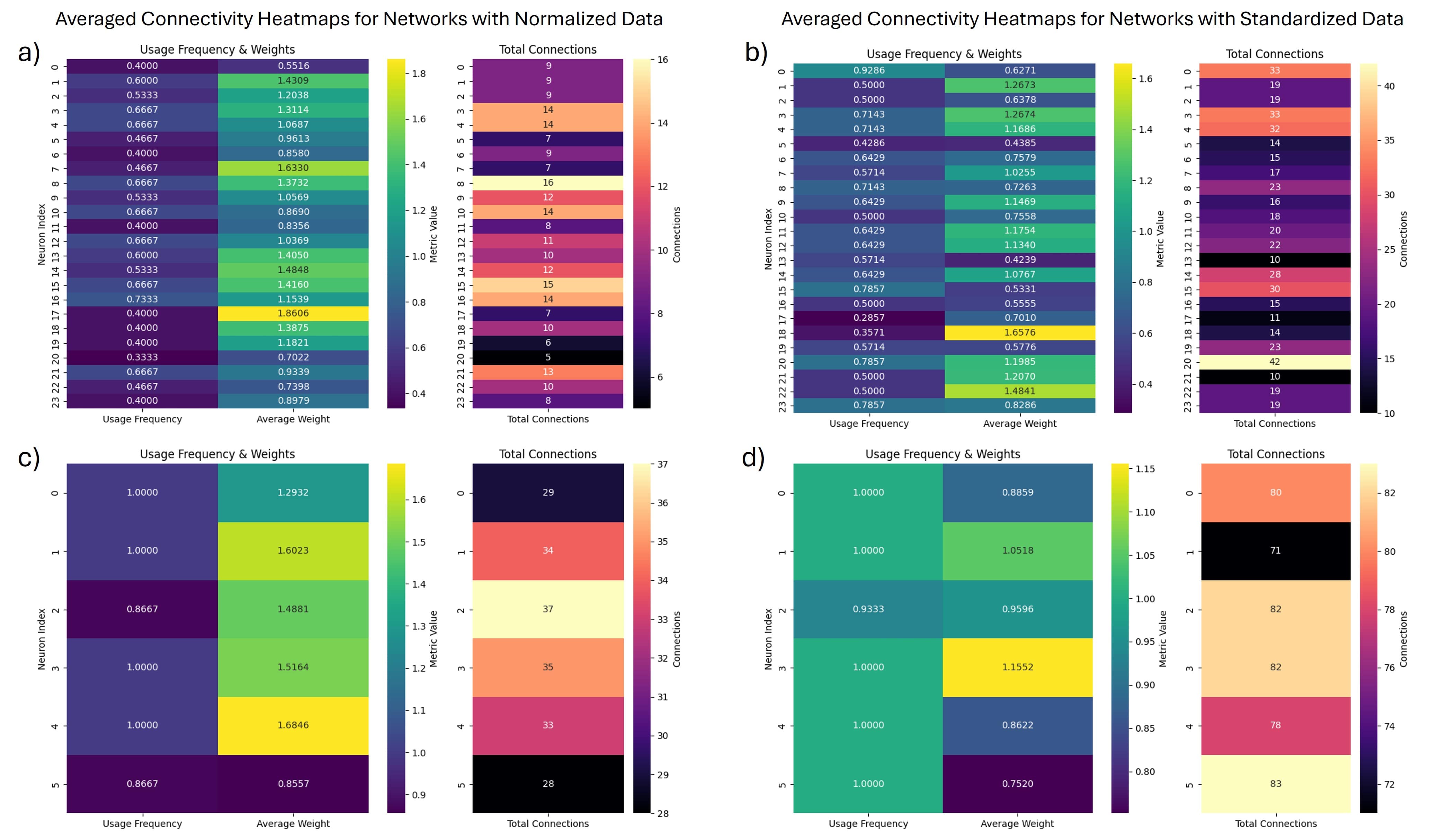}
        \caption{Analysis of the connectivity from input neurons to other neurons in terms of usage frequency, average weights, and total number of connections. The connectivity profiles for 24 and 6 input neurons are summarized for normalized data in (a) and (c), and for standardized data in (b) and (d), respectively, enabling a direct comparison between cases of different input sizes and scaling methods. For each specific configuration, the weights and usage frequencies are averaged across 15 independent runs characterized by identical input--output counts and the same feature scaling protocol. Furthermore, the values for total connections are obtained by summing together the active links developed by the input neurons throughout those 15 runs.}
    \label{fig:heatmaps}
\end{figure}

Our analysis of neuro-evolved architectures revealed a direct link between input preprocessing and network complexity. Networks trained on standardized data developed nearly twice as many active connections as those trained on normalized data, independent of input dimensionality. For instance, in the six-input/two-output configuration, the total number of input-layer connections, across 15 different runs for each type of feature scaling, increased from 188 (normalized) to 406 (standardized). A similar trend was observed for the 24-input case (257 vs. 502 connections).
These findings indicate that standardization promotes richer connectivity patterns (see Figure \ref{fig:heatmaps}), possibly due to broader activation distributions, encouraging NEAT to evolve more complex topologies to capture additional nonlinear dependencies. Despite the increased complexity, standardized networks consistently achieved superior prediction accuracy, suggesting an optimal balance between expressiveness and generalization.

\subsubsection*{Feature Selection and Redundancy Handling}

Beyond architecture discovery, NEAT demonstrated a capacity to act as an implicit feature selector. In experiments using the redundant 24-input configurations, the algorithm systematically emphasized a subset of inputs corresponding to the independent geometric features while suppressing redundant symmetry-derived copies. This behavior was quantified by a composite importance score (Equation \ref{equation}) incorporating input usage frequency, average connection weight, and total connectivity.
\begin{equation}\label{equation}
    S= \text{Usage Frequency} \times |\text{Average Weight}| \times \text{Total Connections}
\end{equation}

\begin{figure}[!tb]
    \centering
    \includegraphics[width=1\linewidth]{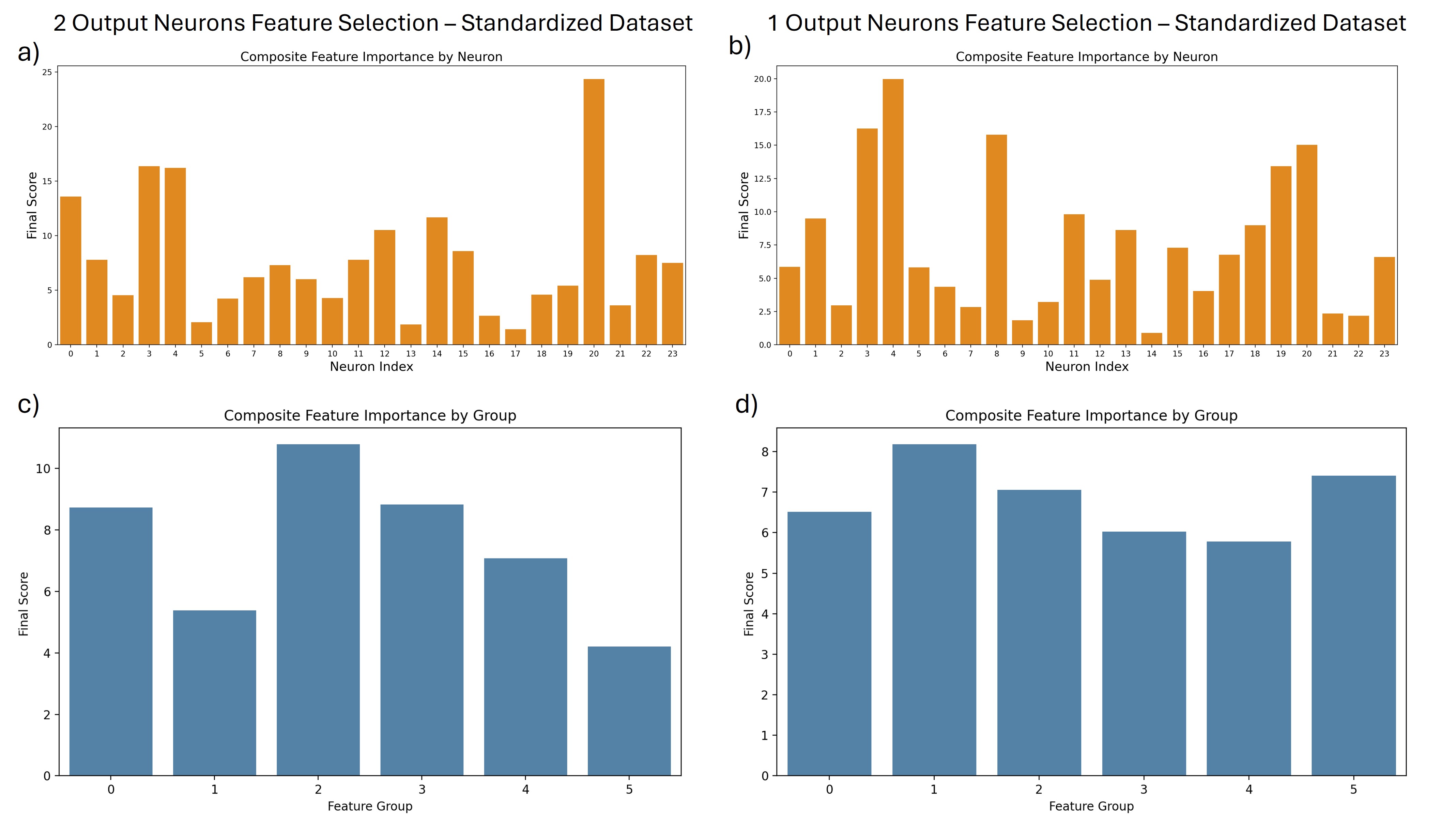}
    \caption{(a), (b) Score assigned to each neuron for the two and one output neuron models, respectively. (c), (d) show corresponding averaged values of the neurons across equivalent coordinates in the four quadrants for these two compared cases. Indices 0 and 1, 2 and 3, and 4 and 5 correspond to the first, second and third coordinate, respectively.}
    \label{fig:Feature Selector}
\end{figure}

Figure \ref{fig:Feature Selector} compares the results for both model types evolved (a and c, b and d, respectively) and examined with standardized data. The feature grouping here (c and d) of the 24-input architectures (as displayed index-wise in a and b) is meant to overlap redundant symmetry-based results in the analysis. 

Inputs corresponding to the second and third coordinate in the base geometry consistently exhibited the highest importance scores across seeds and output configurations, highlighting NEAT's ability to identify the most informative structural descriptors. When multiple outputs were used ($\Delta R_{CD}$ and $R_{pref}$) instead of a single-output model, stronger specialization emerged, with distinct subsets of inputs contributing preferentially to each target, indicative of modular pathway formation during neuroevolution.

According to Fig.~\ref{fig:Feature Selector} c) and d), for each coordinate pair, the x value scored higher than the y one in the two output case, while less distinction is given for the one output case. Markedly, the second coordinate bears the highest accumulated score when optimizing for two optical parameters.

\subsubsection*{Model Deployment in Optimization Pipeline}
The best-performing NEAT-evolved networks were integrated into the full optimization pipeline for  design of GaP/Air metasurfaces with thickness $t = \lambda_0/3$. The resulting designs achieved $\Delta R_{CD}$ values up to 0.0095 and $R_{pref}$ $\approx 0.016$, for the two neuron outputs configuration, and $\Delta R_{CD} \approx 0.0094$ for the one output neuron configuration. These results are comparable to or exceeding those obtained from the previous pipeline runs using manually defined architectures (in this work and in Refs. \cite{mey2022machine, Filippozzi}). These results confirm that neuroevolutionary architecture optimization can maintain or improve optical design performance while eliminating the need for manual network engineering.

\begin{figure}[!tb]
    \centering
    \includegraphics[width=0.75\linewidth]{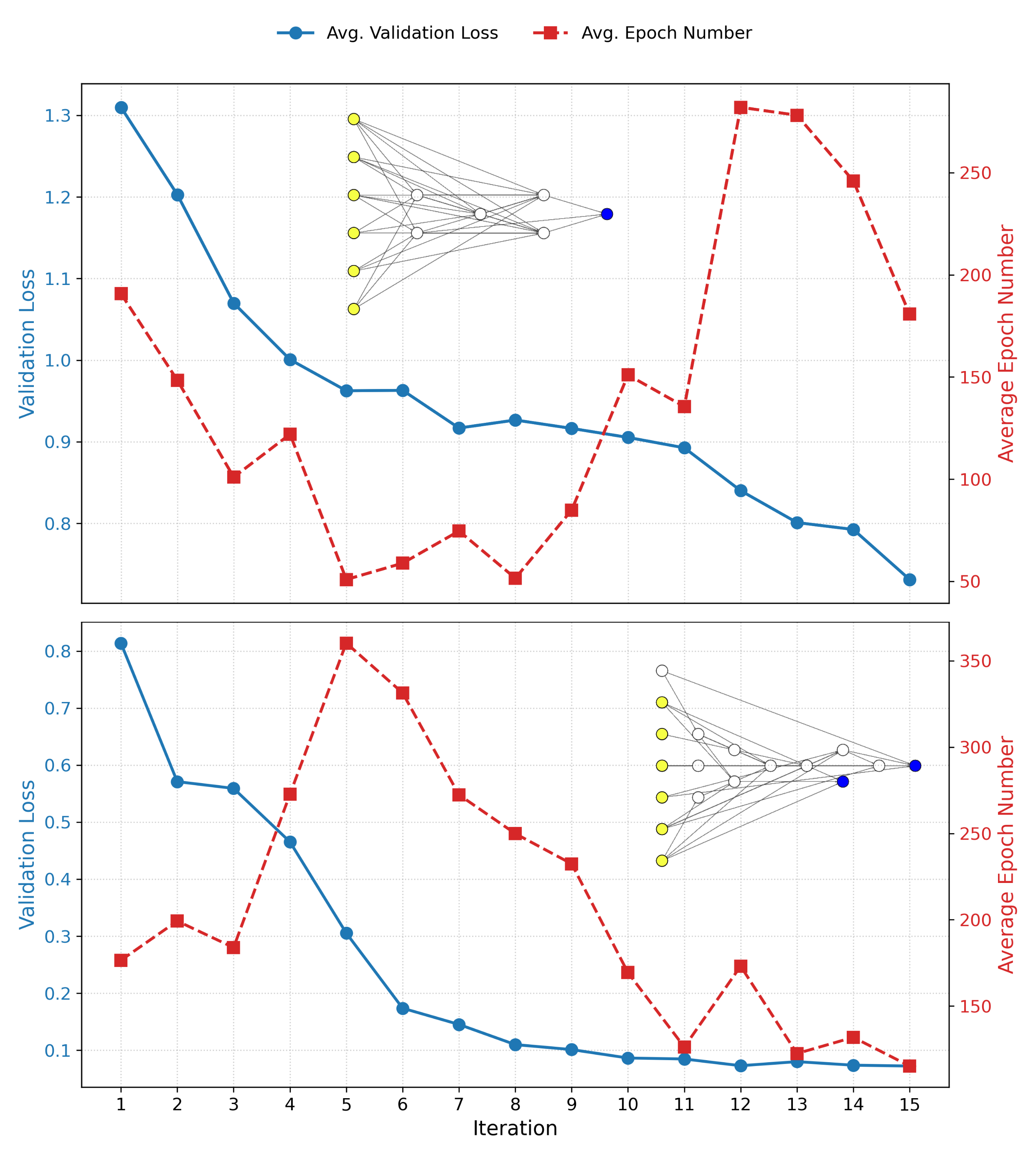}
    \caption{Evolution of validation losses and training durations for each iteration of the NEAT-incorporated optimization pipeline with the one and two output NN models (top and bottom, respectively). MSE average validation losses (8 runs for K-fold cross validation) are displayed alongside the average number of training epochs required for convergence for both cases. Insets: corresponding neuroevolved network architectures utilized as the deep-learning building block for this case.}
    \label{fig:ValLoss}
\end{figure}

As shown on the bottom panel of Fig.~\ref{fig:ValLoss}, the neuroevolved network featuring two output neurons achieved a final MSE validation loss of 0.07 at the conclusion of the training process. This result represents a significant improvement over both the final step of the initial NEAT evolution and the first iteration of the integrated pipeline. In contrast, the configuration consisting of a single output neuron, shown in the top panel of Fig.~\ref{fig:ValLoss}, reached a higher final MSE loss of approximately 0.73. Despite this elevated error metric, the single-output network remained capable of identifying metasurface geometries with exceptionally high $\Delta R_{CD}$ values, successfully fulfilling its design objectives within the optimization framework. Analogously to our previous study reported in Ref.~\cite{Filippozzi}, the training phase utilized the early-stopping principle to automatically adjust the training epochs per pipeline iteration. 

These findings underscore the robustness of the NEAT-evolved models, which balance complexity and predictive accuracy to capture the nonlinear dependencies between structural geometry and chiroptical response.

Overall, NEAT provided an effective means of automating the neural model design process in the optimization of chiral metasurfaces, as an example in the field of metaoptics/metamaterials design improvements. The algorithm successfully adapted to input redundancy, feature scaling, and multiple-objective learning, evolving architectures that can deliver high performance in a machine-learning optimization framework. The consistent advantage of standardized feature scaling suggests that centering the input distribution (while preserving correlations through appropriate rescaling) enhances the search space exploration during neuroevolution.

Advantageously, in our refined deep-learning approach incorporating neuroevolution, three major benefits accompany each other: (i) the data evolution takes place with the help of any used perceptron models through the application of the iterative optimization pipeline (or optionally stochastic evolutionary algorithms), (ii) the structure geometry/design evolution takes place through the (re)training of the underlying neural network models within each pipeline run (reaching more powerful capabilities with better models), and (iii) the machine-learning model's architecture evolution (for that purpose or ``capability boost'') happens through an implemented NEAT algorithm which proposes compact, specialized models for computationally-efficient prediction in the given optimization scenario. 

This comprehensive framework, promises additional benefits in conjunction with data augmentation and network finetuning with experimental data, offering productive design optimization for different use cases and with transfer learning capacity. Additionally, the coevolution of the network model and (coevolution of) production samples with feedback to the model framework through (instantaneous) measurement data can lead to dynamic, self-driven fabrication pipelines with self-adjustment, as well as automation, of design and production capabilities. Employing the pipeline not only contributes to geometry evolution (i.e. design suggestions within the so far accessed solution space), but inherently enriches the available dataset with newly-simulated, high-performance structures for future examinations and model evolutions. Eventually, the role of the model architect being transferred from humans to machines---with the capability of evolving models in a problem specific manner and adapting to underlying physics efficiently---could critically enable future data-driven workflows, which would utilize machine-learning building blocks within autonomous labs, to become fully controlled and carried out through agentic AI frameworks.

\section{Conclusion}

In this work we demonstrated the successful implementation of the NeuroEvolution of Augmenting Topologies (NEAT) algorithm for a deep-learning optimization of chiral dielectric metasurfaces. By allowing the neural architecture to evolve autonomously, NEAT eliminates the need for manual network design and provides a data-driven approach to model adaptation. 

The neuroevolutionary process promotes a balance in model complexity and performance, producing topologies that adaptively captured nonlinear dependencies between structural geometry and chiroptical response. Moreover, NEAT acted as an implicit feature selector, efficiently filtering redundant symmetric parameters and emphasizing the most informative geometric descriptors. Particularly when trained on standardized datasets, NEAT-evolved models reached better accuracy and generalization, ultimately outperforming fixed-topology counterparts after further finetuning of weights inside our training and inference pipeline.  Deployed in the optimization framework, the neuroevolved networks enabled the design of GaP/Air metasurfaces with circular dichroism and reflectance properties comparable to, or better than, those obtained with manually designed NN architectures with 2 hidden layers and an overall higher number of hidden neurons.

Overall, this work establishes NEAT-driven neuroevolution as a powerful extension to machine learning–based nanophotonic design pipelines. The proposed approach enables fully automated, self-adjusting optimization workflows that can be generalized to other metasurface platforms and extended to multi-objective or dynamic design tasks in optical, electronic, and quantum photonic systems.
Looking forward, the paradigm of a self-developing optimization pipeline could be further enhanced through the integration of agentic AI. In such a framework, agentic systems could autonomously design and develop sophisticated optimization procedures independently, further reducing human intervention toward truly self-configuring photonic design environments.

\section*{Acknowledgments}
Financial support by the Deutsche Forschungsgemeinschaft (German Research Foundation, Grant No.: DFG RA2841/12-1)---Projektnummer 456700276 (Project ID)---is acknowledged. The authors
thank O. Mey for the development of the initial reinforced supervised learning algorithm.

\bibliographystyle{unsrt}
\bibliography{bib.bib}

\appendix

\section{NEAT Algorithm Evaluation in 1-Output Configuration}
\label{appendix:NEAT_1_Neuron}

\begin{figure}[!ht]
    \centering
    \includegraphics[width=1.0\linewidth]{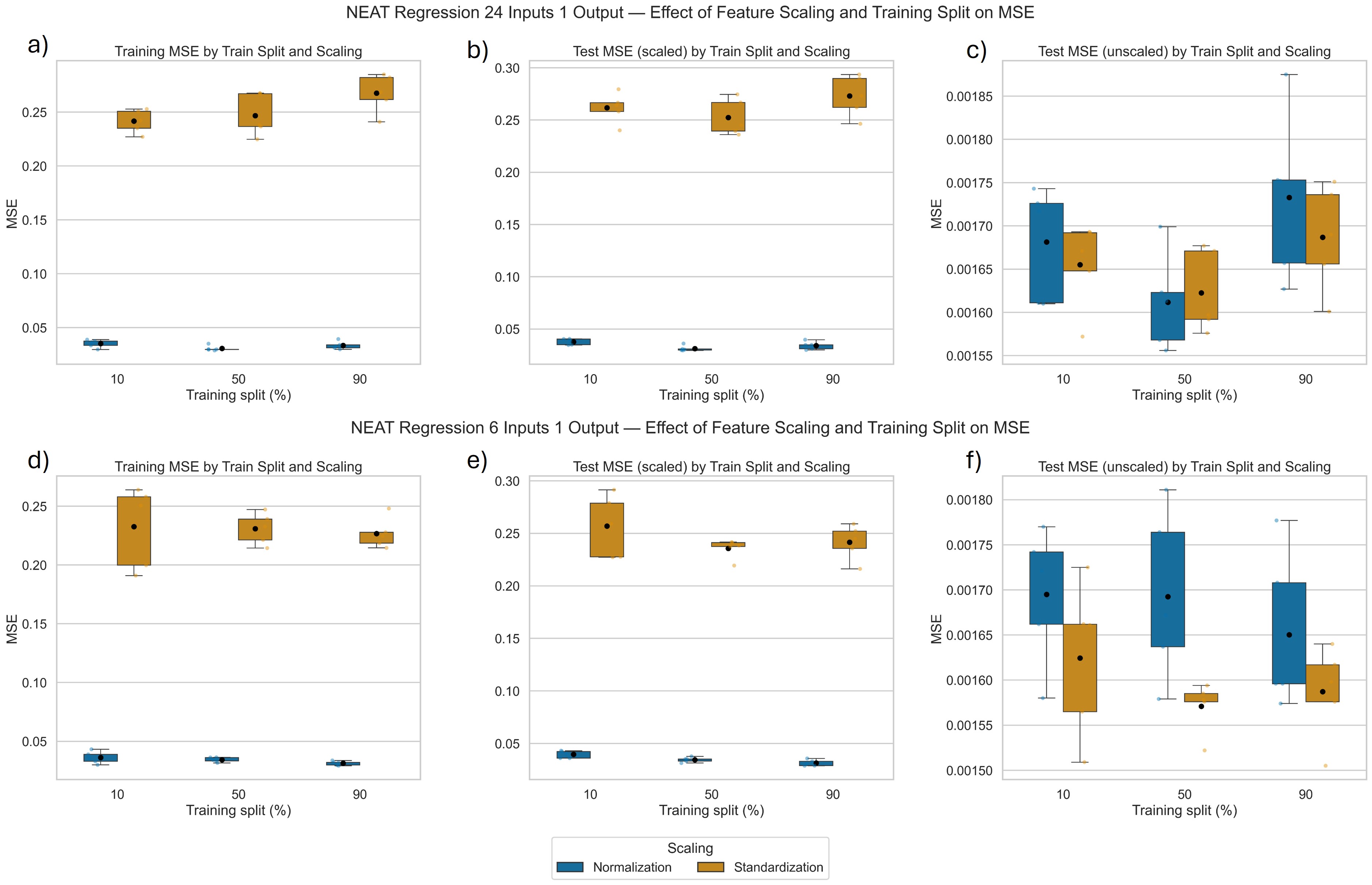}
    \caption{Overview of the NEAT algorithm results obtained for 1 output neuron, with 24 and 6 input neurons used in the model (top and bottom, respectively). a) and d) show the averaged training losses for every train-test split of the dataset in a comparison of normalized (dark colored) and standardized (light colored) data. Error bars obtained from averaging over 5 repetitions are provided for each value. b) and e) similarly show the
    respective averaged test losses. c) and f) summarize the re-scaled results of the
    tests, that is, undoing the "scaler". The direct comparison displays lower MSE values in the case of less input features and standardization for sufficiently sized training data.}
    \label{fig:NEAT 1 Neurons}
\end{figure}

Figure~\ref{fig:NEAT 1 Neurons} summarizes the averaged NEAT results for the 24 inputs/1 output, and 6 inputs/1 output configuration, analog to Fig.~\ref{fig:MSE}.
From these results, it can be stated that when 6 inputs are used, the feature scaling has a significant effect on the prediction capabilities of the evolved networks, with standardization the method providing superior results, while the dataset size contributed to the outcomes to a certain extent as well. However, in the case of 24 input neurons, mainly the dataset size mattered, with a dataset size of 4800 training examples (50:50 split) leading to improved results, whereas feature scaling did not play a significant role.

\section{Run-time comparisons for the NEAT algorithm}
\label{appendix:run-time}

\begin{figure}[!ht]
    \centering
    \includegraphics[width=1\linewidth]{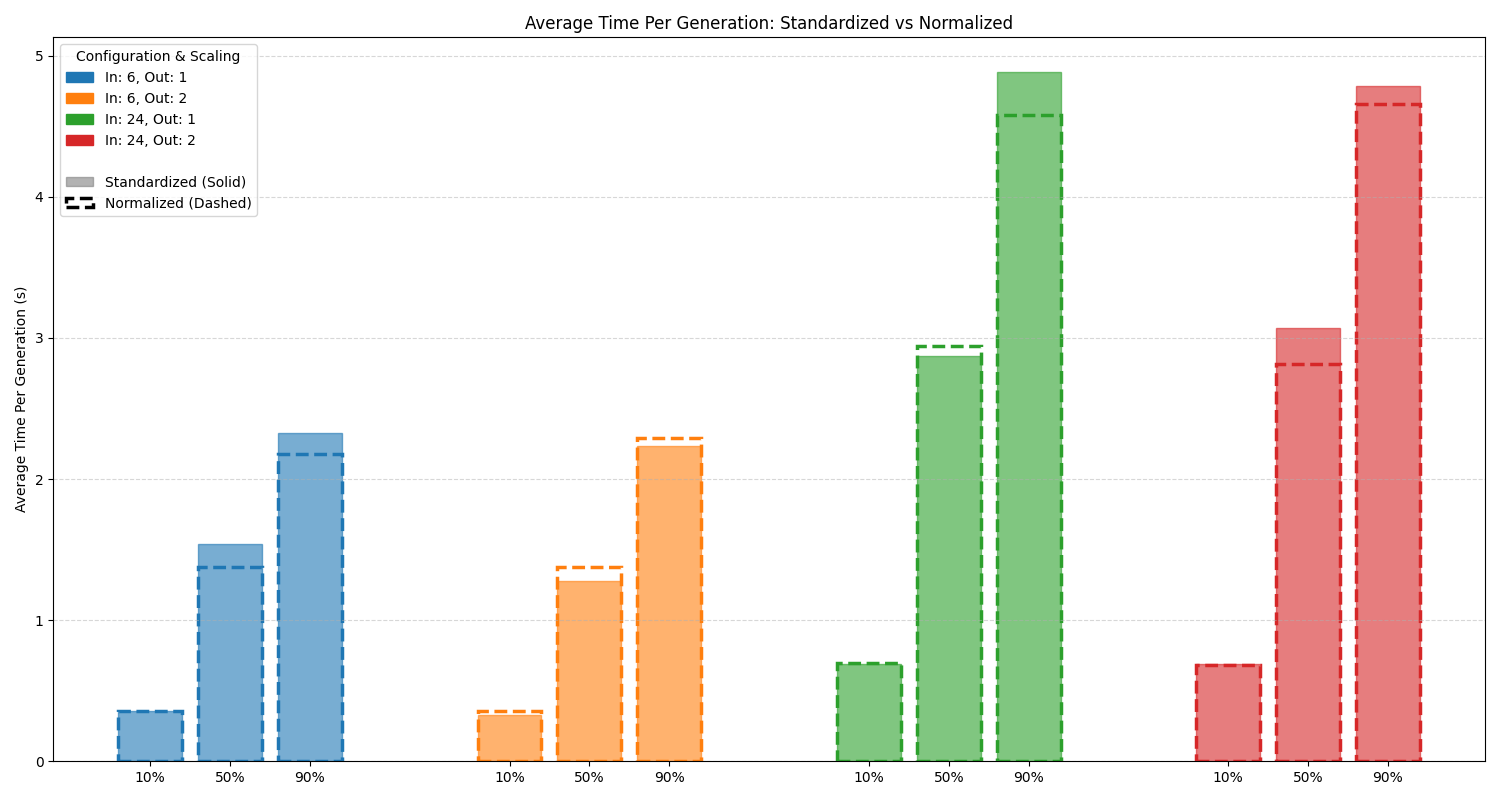}
    \caption{Summary of NEAT computational performance across varying network architectures and training set sizes. Results are averaged over five independent trials ($n=5$). The input dimensionality corresponds to "One Quadrant" (6 neurons) and "All Quadrants" (24 neurons) configurations, while the output layers consist of either 1 or 2 neurons. Training percentages represent the fraction of the total dataset utilized during the evolution process. Each trial has 2000 generations.}
    \label{fig:average_time}
\end{figure}

To provide an overview of run-times during the NEAT-based model optimization runs, Fig.~\ref{fig:average_time} summarizes computational performances for the different input/output dimensionalities and for the two applied feature scalings.

\section{Hyperparameters and Problem Configuration}
\label{appendix:hyperparameters}

This appendix details the exact configuration for the NEAT implementation used in this study, utilizing the \texttt{TensorNEAT} library based on JAX.

\begin{table}[!ht]
    \centering
    \caption{Problem-Specific Configuration}
    \label{tab:problem_config}
    \begin{tabular}{l c l}
        \hline
        \textbf{Parameter} & \textbf{Value} & \textbf{Description} \\
        \hline
        Input Nodes & 6, 24 & Fixed amount of input nodes \\
        Output Nodes & 1, 2 & Fixed amount of output nodes \\
        Fitness Function & Negative MSE & Minimizing Mean Squared Error \\
        Data Normalization & Standard Scaler, MinMax & Standardization and Normalization\\
        Random Seed & [42, 90, 231, 656, 2341] & Reproducibility seed  \\
        \hline
    \end{tabular}
\end{table}

\subsection{NEAT Algorithm Hyperparameters}

The problem-specific configuration and hyperparameters summarized in Tables~\ref{tab:problem_config} and \ref{tab:neat_hyperparameters}, respectively, define the evolutionary process.

\begin{table}[!ht]
    \centering
    \caption{NEAT Algorithm Hyperparameters. $^{(1)}$Max nodes was set to 50 for the tests with 6 input neurons, and to 60 for the tests with 24 input neurons.
    $^{(2)}$ReLU and LeLU were used along with Standardization, Tanh and Sigmoid were used along with Normalization.}
    \label{tab:neat_hyperparameters}
    \begin{tabular}{l c}
        \hline
        \textbf{Parameter} & \textbf{Value} \\
        \hline
        \multicolumn{2}{c}{\textbf{General Settings}} \\
        \hline
        Population Size & 1500 \\
        Generation Limit & 2000 \\
        Max Nodes & 50, 60$^{(1)}$ \\
        Max Connections & 100 \\
        \hline
        \multicolumn{2}{c}{\textbf{Genome Structural Mutation}} \\
        \hline
        Connection Add Prob. & 0.2 \\
        Connection Delete Prob. & 0.2 \\
        Node Add Prob. & 0.02 \\
        Node Delete Prob. & 0.02 \\
        \hline
        \multicolumn{2}{c}{\textbf{Weight \& Bias Mutation}} \\
        \hline
        Weight Mutation Rate & 0.5 \\
        Weight Mutation Power & 0.15 \\
        Weight Range & $[-5.0, 5.0]$ \\
        Bias Mutation Rate & 0.2 \\
        Bias Mutation Power & 0.15 \\
        Bias Range & $[-5.0, 5.0]$ \\
        \hline
        \multicolumn{2}{c}{\textbf{Speciation}} \\
        \hline
        Compatibility Threshold  & 2.0 \\
        Disjoint Coefficient  & 1.0 \\
        Weight Coefficient  & 0.4 \\
        Survival Threshold & 0.1 \\
        Max Stagnation & 15 \\
        \hline
        \multicolumn{2}{c}{\textbf{Network Properties}} \\
        \hline
        Hidden Activations & [ReLU, LeLU], [Tanh, Sigmoid]$^{(2)}$ \\
        Output Activation & Identity \\
        Aggregation Method & Sum \\
        \hline
    \end{tabular}
\end{table}

\end{document}